\documentstyle[preprint,aps]{revtex}
\begin{document}

\input epsf

\draft

\title{The $Q^2$ dependence of the hard diffractive photoproduction
of vector meson or photon and the range of pQCD validity}
\author{I.F.~Ginzburg\thanks{e-mail: ginzburg@math.nsk.su},
D.Yu.~Ivanov\thanks{e-mail: d-ivanov@math.nsk.su}}
\address{Institute of Mathematics, 630090 Novosibirsk, Russia}
\date{\today}

\maketitle

\begin{abstract}
We consider two coupled problems.

We study the dependence on photon virtuality $Q^2$ for the
semihard quasi--elastic photoproduction of neutral vector mesons
on a quark, gluon or real photon (at $s\gg p_{\bot}^2,\;Q^2; \;
p_{\bot}^2\gg \mu^2 \approx (0.3$ GeV)$^2$). To this end we
calculate the corresponding amplitudes (in an analytical form)
in the lowest nontrivial approximation of perturbative QCD.  It
is shown that the amplitude for the production of light meson
varies very rapidly with the photon virtuality near $Q^2=0$.

We estimate the bound of the pQCD validity region for such
processes. For the real incident photon the obtained bound for
the $\rho$ meson production is very high. This bound decreases
fast with the increase of $Q^2$, and we expect that the virtual
photoproduction at HERA gives opportunity to test the pQCD
results. The signature of this region is discussed.  For the
hard Compton effect the pQCD should work good at not too high
$p_{\bot}$, and this effect seems measurable at HERA.
\end{abstract}


\section{Introduction}

The diffractive photoproduction of neutral vector mesons or
photon was investigated in many theoretical [1--16] and
experimental [17--19] papers.

Below we study the photoproduction of light vector meson or
photon (hard Compton effect) on a quark or gluon, initiated by
both real and virtual photon $\gamma^*$ (having virtuality
$Q^2$):
\begin{equation}
\gamma^* q\to Vq,\quad \gamma ^* g\to Vg\quad  (V=\rho ,\phi ,...,
\gamma ,... )\label{0}
\end{equation}
in the region of parameters where the perturbative QCD (pQCD)
validity is beyond doubts:
\begin{equation}
s\gg p_{\bot}^2,\;Q^2;\;\;\; p_{\bot}^2\gg\mu^2\qquad (\mu\approx
0.2\div 0.3 \mbox { GeV}).
\label{range}
\end{equation}
The transverse momentum of produced meson relative to collision
axis $\mbox{$p_{\bot}\,$}$ is small as compared to the energy but
it is large as compared to the QCD scale $\mu \approx 0.3$ GeV.

Besides, we consider the similar photoproduction of heavy vector
mesons on a quark, gluon or other photon:
\begin{equation}
\gamma^*q\to \tilde{V}q,\;\;\gamma^* g\to\tilde{V}g,\;\;
\gamma^*\gamma \to \tilde{V}\tilde{V}^{\prime}\quad (\tilde{V},
\tilde{V}'=J/\Psi,...).\label{0a}
\end{equation}

The photoproduction on quarks and qluons can be studied as that
on a proton in the events with the rapidity gap $\eta>\eta_0$
between produced meson and other produced hadrons X.  The
photon--photon processes can be studied at the future photon
colliders \cite{GKST}.

The cross section of the process $\gamma^* p \to V X$ with
rapidity gap is related with that for the photoproduction on a
quark and a gluon via the gluon and quark densities in a proton
$G(x,t)$ and $q(x,t)$:
\begin{eqnarray}
\frac{d^2\sigma (\gamma^* p\to VX)}{dtdx}&=&
\sum_f\left(q(x,t)+\bar q(x,t)\right)
\frac{d\sigma(\gamma^* q\to Vq)}{dt}+\nonumber\\
&&G(x,t)\frac{d\sigma(\gamma^* G\to VG)}{dt};\quad
x>\frac{4 p_{\bot}^2}{s}\cosh^2\frac{\eta_0}{2}.\label{strf}
\end{eqnarray}

The main object, studied in these processes, is called the
perturbative Pomeron (pP). Its theoretical and experimental
study is of great interest, since this object should be common
for different reactions and it is sensitive to the inner
structure of pQCD. Only the results in the Leading Log
Approximation (LLA) are known until now, that is the BFKL
Pomeron \cite{Lip}.

To estimate the bounds of the pQCD validity region, we simulate
the nonperturbative effects near these bounds by the specific
model. The idea of this model is to use the pQCD equations, in which
quark mass is considered as a parameter (which is near the
constituent quark mass). We also use this model for the
qualitative description of some phenomena outside this bound.

Our efforts are focused on the problems: {\em What are the main
features of $Q^2$-dependence in these processes within pQCD,
without any phenomenological hypotheses?  What are the bounds of
pQCD validity at the description of diffractive processes?} To
this end we restrict our consideration by the calculation in the
lowest nontrivial approximation of pQCD --- two--gluon exchange
in the $t$-channel (see Fig. \ref{fig1}). We obtain that the
cross sections of light meson photoproduction decrease very fast
with virtuality near mass shell. Besides, for these processes
and for hard Compton effect with real incident photon the
influence of nonperturbative effects is very essential in the
very large interval of $p_{\bot}$, the pQCD become valid at very
large values of $p_{\bot}$ only. We present some arguments, why
these features could be reproduced in the entire amplitude
(beyond two--gluon approximation). We give the definite
signature for the pQCD validity in the photoproduction of light
mesons, which is independent on the validity of two--gluon
approximation.

In this respect, the calculations with heavy mesons are
presented mainly for the comparison to make clearer some
details. Both our calculations here and the comparison of LLA
results \cite{ForRys,Iv} show that the naive extension of the
results obtained for heavy mesons to the photoproduction of
light mesons or photons is dangerous procedure.

Additionally, the obtained results provide the opportunity to
discuss a relation between the point--like and the hadron--like
components of a photon.

The known for us papers, treated the similar problems, are
discussed in the end of paper.

The study of pQCD validity in the discussed processes is on line
with that in the other exclusive reactions. The relation between
perturbative and nonperturbative contributions to the amplitudes
of exclusive processes was discussed widely (see refs.
\cite{I,Br} and references therein). The advantage of processes
considered is the possibility to study this subject by the two
probes simultaneously --- via investigation of the dependence on
both the produced meson transverse momentum $p_{\bot}$ and the
photon virtuality $Q^2$.

\section{Basic relations}

The process discussed can be described as two stage one. At the
first stage, a photon decays into a $q\bar q$ pair, the
quarks with energies $\varepsilon_i$ move along the photon
momentum.  Their transverse momenta are relatively small and the
total energy of quark's pair is close to the energy of the
photon, $\varepsilon_1+\varepsilon_2\approx E$. This first stage
describes also the processes with the production of jet--like
hadron system (both resolved for the two quark jets and
unresolved one) with the rapidity gap.

At the second stage the quarks are glued into meson.

The basic kinematical notations are presented in Fig. \ref{fig1}.
We denote also the virtuality of photon by $Q^2\equiv -p_{1}^2
>0$, the quark mass --- by $m$, the transverse momentum of
produced meson (relative to the collision axis) --- by ${\bf
p_{\bot}}$.  For the description of the photon fragmentation into
quarks, we use quark spinors $u_1=u(q_1)$ and $u_2=u(-q_2)$. The
relative motion of the quark and antiquark is described by the
variable $\xi $:
$$
\xi =\frac{2\left(q_1-q_2\right)p_2}{s}\equiv \frac
{\varepsilon _1-\varepsilon _2}{E} ;\qquad -1\leq \xi \leq 1.
$$
Next, we denote
\begin{equation}
{\bf n} = {{\bf p}_{\bot} \over \mid {\bf p}_{\bot}\mid};\quad
\delta = {2m\over p_{\bot}};\;\; u={Q^2\over {\bf p}_{\bot}^2};\;
\; v= \delta^2 + (1-\xi^2)u= {4m^2+(1-\xi^2)Q^2 \over {\bf p}_{\bot}^2}.
\label{u}
\end{equation}
Besides, $e= (0,{\bf e},0)$ and $e_V=(0,{\bf e}_V,0)$ are the
polarization vectors of the transverse photon and the
transversely polarized vector meson.

As usually, $ \alpha _s= g^2/4\pi,\;\alpha =e^2/4\pi =1/137,\;Q_q
e$ is the quark charge, $N=3$ is the number of colors.
\vspace{0.3cm}

{\bf Impact representation}
\vspace{0.3cm}

The amplitude of the process in the lowest nontrivial pQCD order
is described by the diagrams of Fig. \ref{fig1} (with the
accuracy $\sim p_{\bot}^2/s,\; Q^2/s $). Just as in refs.
\cite{GPS1,GPS2,GIv}, the sum of these diagrams is transformed
with the same accuracy to the integral over the gluon transverse
momentum --- {\em the impact representation}:
\begin{equation}
M_{\gamma ^*q\to Vq}=is\int \frac {J_{\gamma ^*V}({\bf
k}_{\bot},{\bf p}_{\bot})\; J_{qq}(-{\bf k}_{\bot},-{\bf
p}_{\bot})} {{\bf k}^2_{\bot}\; ({\bf k}_{\bot}- {\bf
p}_{\bot})^2}\frac{d^2k_{\bot}}{(2\pi)^2}.\label{4}
\end{equation}

Impact--factors $J_{\gamma ^* V}$ and $J_{qq}$ correspond to the
upper and the lower blocks in Fig. \ref{fig1}. They are
$s$--independent. The entire dependence on the photon virtuality
is concentrated in the impact--factor $J_{\gamma^* V}$. For
colorless exchange the impact--factors contain factors
$\delta_{ab}$, where $a$ and $b$ are the color indices of the
exchanged gluons.

The impact--factors, which describe the transition between two
colorless states vanish when the gluon momenta tend to zero:
\cite{GPS1}:
\begin{equation}
J_{\gamma^* V}({\bf k}_{\bot},{\bf p}_{\bot})\to 0\;\mbox{ at }\;
\left\{\begin{array}{cr}
{\bf k}_{\bot}&\to 0,\\ {\bf (p-k)}_{\bot}& \to 0.
\end{array}\right.
\label{dipol}
\end{equation}

This general property ("dipole shielding" or "quark coherence")
takes place independent on the validity of perturbation
expansion. In the coordinate space this property can be treated
as zero's color charge of an object.

The derivation of impact representation and impact--factors
repeats in the main features that given in Appendices to ref.
\cite{GPS1} (see refs. \cite{LFr},\cite{ChWu} also) with two
variations.

First, one should consider separately the impact--factors of
a photon to meson transitions for the transverse ($\gamma^*_T$)
and scalar (or longitudinal) ($\gamma^*_S$) initial off--shell
photons.

Second, Sudakov variables are introduced for the reaction with
"massive" collided particles. All momenta are decomposed over the
plane perpendicular to the reaction axis and over the light-like
four--vectors combined from the initial ones: $p'_1 = p_1 -
(p^2_1/s) p_2=p_1 +(Q^2/s) p_2\;$, $\;p'_2=p_2 -(p^2_2/ s) p_1 $.
It leads to the single replacement
\begin{equation}
m^2\to m^2+ Q^2(1-\xi^2)/4.\label{mq}
\end{equation}
in the denominators of quark propagators.

{\bf Relation to the BFKL Pomeron.}
\vspace{0.3cm}

Let us discuss briefly the relation with the entire pQCD series
in LLA (the BFKL Pomeron \cite{Lip}). The modern using of BFKL
for the diffractive processes at high $p_{\bot}$ was initiated
by the paper \cite{MT}.

In the LLA the impact representation transforms to the form (see
Fig.\ref{fig2}):
\begin{equation}
M_{\gamma ^*q\to Vq}=is\int J_{\gamma ^*V}({\bf k}_{\bot},{\bf
p}_{\bot})\; J_{qq}(-{\bf k'}_{\bot},-{\bf p}_{\bot}) {\cal P}
(s;{\bf p_{\bot}, k_{\bot},k'_{\bot}})
\frac{d^2 k_{\bot}d^2 k'_{\bot}}{(2\pi)^4}. \label{pomerimp}
\end{equation}

The discussed lowest nontrivial approximation of pQCD (\ref{4})
corresponds to
$$
{\cal P}(s;{\bf p_{\bot},k_{\bot},k'_{\bot}})=
\frac{(2\pi)^2\delta({\bf k}_{\bot}-{\bf k}'_{\bot})}
{{\bf k}^2_{\bot}\; ({\bf k}_{\bot}- {\bf p}_{\bot})^2}
\delta_{aa'}\delta_{bb'}.
$$

The kernel ${\cal P}$ of eq. (\ref{pomerimp}) corresponds to the
LLA for the pP.

In accordance with predictions of refs. \cite{GPS1,GPS2}, the
amplitude (\ref{pomerimp}) in its asymptotic form (the high
energy asymptotic of LLA result) is Regge--like:
\begin{equation}
M=is G_{\gamma^* V} (\mbox{$p_{\bot}\,$},Q^2)\cdot
K(s/\mbox{$p_{\bot}^2\,$})\cdot G_{q q}(p_{\bot}).\nonumber
\end{equation}
At $ p_{\bot}^2\gtrsim Q^2$ the LLA is governed by the unique
large logarithm $\ln(s/p_{\bot}^2)$, and the lowest nontrivial
approximation for the impact--factors is valid.  In this case
the vertex $G_{\gamma^* V}$ is the convolution of our
$\gamma^*\to V$ impact--factor with some factor from ${\cal P}$.
The corresponding integration is similar to that in our case,
and the $Q^2$ dependence near mass shell is expected to be
roughly the same.

At $Q^2\gg p_{\bot}^2$ the additional large logarithm $\ln(Q^2/
p_{\bot}^2)$ appear. In this case the (unknown now) higher order
corrections should be taken into account even at the description
of the impact--factors. It is the first reason, why the $Q^2$
dependence in this region cannot be predicted definitely now.

The kernel $K$ is pP itself, it is obtained in refs. \cite{Lip},
one can write for $\rho\gg 1$
\begin{equation}
K=\frac{e^{\rho \ln4}}{(7\zeta(3)\rho)^{3/2}};\quad
\rho=\frac{6\alpha_s(c_1p_{\bot}^2)}
{\pi}\ln{s\over c_2p_{\bot}^2}.\label{BFKLpom}
\end{equation}
The quantities $x_i$ here cannot be determined within LLA (even at
$Q^2\ll p_{\bot}^2$ and for light mesons or photons).  At $Q^2\gg
p_{\bot}^2$ these coefficients are completely unknown
\cite{Lipatov}. It is the second reason, why the $Q^2$
dependence in this region cannot be predicted definitely now.

The real influence of BFKL corrections is different for
different processes even for real incident photons (due to
important numerical factors). For the $J/\Psi$ production the
BFKL Pomeron enhances the two--gluon result already at not too
small $\rho>0.8$ and pure two--gluon approximation practically
lacks range of validity \cite{ForRys}. On the other hand, the
Regge--like contribution (\ref{BFKLpom}) is larger than the
two--gluon result only at $\rho>3.4$ for the $\rho$ production
with real photon \cite{Iv} and at $\rho\sim 1$ for the hard
Compton effect ($\gamma q \to \gamma q$) \cite{Iv1}.  At smaller
$\rho$ values the calculations in the discussed two--gluon
approximation seems more adequate for the description of data.

\section{ Impact--factors}

{\bf Quark and gluon impact--factors}
\vspace{0.3cm}

The impact--factors of $q\to q$ (from \cite{GPS1}) and $g\to g$
transitions through the colorless exchange are similar:
\begin{equation}
J_{qq}=g^2\;{\delta_{ab} \over 2N};\quad J_{gg}=-g^2\delta _{ab}
\;{N\over N^2-1}.\label{14}
\end{equation}

The helicity and color state of the quark or gluon target are
conserved in these vertices.

The relations (\ref{14}) show that the cross section for the
photoproduction of vector meson on a gluon is about 5 times
larger than that on a quark:
\begin{equation}
d\sigma _{\gamma ^* g \to Vg} = \left(\frac{2N^2}{N^2-1}\right)^2
d\sigma _{\gamma ^*q \to Vq}={81\over 16 } d\sigma _{\gamma ^* q
\to Vq}.\label{GlQ}
\end{equation}
It means that the photoproduction of vector meson on a proton
with the rapidity gap can be used for the study of the gluon
content of a proton.

Besides, this relation allows us to present below the formulae
for the photoproduction on a quark only.

\vspace{0.3cm}

{\bf Impact--factor $J_{\gamma^*\to q\bar q}$}
\vspace{0.3cm}

{\large\em The impact--factor for a transverse photon} has the
same form as for the on shell photon \cite{GPS1} but with the
replacement (\ref{mq}) in denominators:
\begin{equation}
J_{\gamma ^*_T\to q\bar q}=eQ_qg^2 \;{\delta_{ab}\over 2N}\;{\bar
u}_1\left[mR(m){\hat e}-\left(1+\xi\right){\bf P}(m){\bf e}-\hat
P(m)\hat e \right]  \; {{\hat p}_2 \over s}\;u_2. \label{5}
\end{equation}

Here transverse vector $P(m)=(0,{\bf P}(m),0)$ and scalar $R(m)$
are:
\begin{eqnarray}
{\bf P}(m)&=&
\left[\frac{{\bf q}_{1\bot}}{ {\bf q}^2_{1\bot}+m^2+(1-\xi
^2)Q^2/4} \;+ \;\frac {{\bf k}_{\bot}-{\bf q}_{1\bot}}{ ({\bf
k}_{\bot}-{\bf q}_{1\bot})^2 +m^2+(1-\xi ^2)Q^2/4}\right] -
\label{7}\\
&&-\left[{\bf q}_{1\bot}\leftrightarrow {\bf
q}_{2\bot}\right];\nonumber\\ &&\nonumber\\ R(m)&=&\left[{1 \over
{\bf q}^2_{1\bot}+m^2+(1-\xi ^2)Q^2/4} \;- \; {1 \over ({\bf
k}_{\bot}-{\bf q}_{1\bot})^2 +m^2+(1-\xi ^2)Q^2/4}\right]
+\label{7a}\\ && +\left[{\bf q}_{1\bot}\leftrightarrow {\bf
q}_{2\bot}\right]. \nonumber
\end{eqnarray}

To describe {\large \em the impact--factor for a scalar photon},
it is necessary to define the polarization vector of scalar
photon $e_S$. Taking into account the gauge invariance, in our
kinematical region one can use a reduced form of this vector $
e_S=2\sqrt{Q^2}\; (p'_2/ s)$. Then the calculations similar to
those for $T$ photon:
\begin{equation}
J_{\gamma ^*_S q\bar q}=-eQ_qg^2{
\delta_{ab}\over 2N}\;  \frac{1-\xi ^2}{2}\sqrt{Q^2}\;R(m)\;
{\bar u}_1 \; {{\hat p}_2 \over s}\; u_2.
\label{6}
\end{equation}

It is easily seen that these impact--factors obey eq.
(\ref{dipol}).
\vspace{0.3cm}

{\bf Impact--factors for a meson production}
\vspace{0.3cm}

To produce a meson, the relative transverse momenta of quarks
should be small ($\stackrel {<}{\sim} \mu$). With our accuracy
($\mu^2/{\bf p_{\bot}^2}\ll 1$) the transverse momenta of
quarks relative to collision axis are proportional to their
energies $\varepsilon_i$, i.e.
$$
{\bf q}_{1\bot}=\frac{1}{2}(1+\xi ){\bf p}_{\bot},\quad {\bf
q}_{2\bot}=\frac{1}{2}(1-\xi ){\bf p}_{\bot},\quad
\varepsilon_{1,2}=\frac{1}{2}(1\pm \xi )E.
$$

The $q \bar q \to V$ transition is described, as usual (see
\cite{LeBr}), by the change of product $\bar u_1...u_2$ for
meson wave function $\varphi_V (\xi)$:
\begin{equation}
Q_q\bar u_1\dots u_2\to \frac{Q_V}{4N}\int \limits_{-1}^{1}d\xi
\left\{\begin{array}{ccc}
f_V^L \varphi_V^L(\xi )\mbox {Tr}\left( \dots\hat p_3\right)&
\mbox{ for }V_L\\
f_V^T \varphi_V^T(\xi ) \mbox {Tr}\left(\dots
\hat e^*_V\hat p_3\right) & \mbox{ for } V_T.
\end{array}\right.\label{2}
\end{equation}
(The trace over vector and color indices is assumed). The
quantity $Q_V$ relates to the quark charges in the meson $V$.
The specific form for these wave functions is given in eq.
\ref{3a},
eq. \ref{3}. We use the coupling constants from refs.
\cite{ChAZh},\cite{BaGr}, see Table \ref{tab1}.

The impact--factors $J_{\gamma^* V}$ are obtained by the
substitution of eq. (\ref{2}) into eqs. (\ref{5}),(\ref{6}):

a) For {\em a transverse photon} we have two opportunities:
\begin{equation}
J_{\gamma _T ^* V}({\bf k}_{\bot},{\bf p}_{\bot})= {1\over 2}eQ_V
g^2 {\delta _{ab} \over 2N}\; \int
\limits^{1}_{-1}d\xi \left\{
\begin{array}{ll}
(-f^L_V)\varphi ^L_V(\xi )\;\xi \;({\bf P} {\bf e}) & \mbox{ for
} V_L\\ f^T_V\varphi ^T_V(\xi )\;mR\;({\bf e}{\bf e}^*_V) & \mbox
{ for } V_T.\end{array}\right.\nonumber
\end{equation}

b) {\em A scalar photon} produces a longitudinal vector meson
only:
\begin{equation}
J_{\gamma _S ^* V _L}({\bf k}_{\bot},{\bf p}_{\bot})= -{1\over
2}eQ_V g^2 {\delta _{ab} \over 2N}\; \int
\limits^{1}_{-1}d\xi f^L_V \varphi ^L_V (\xi)
{1-\xi ^2 \over 2}\; \sqrt{Q^2}\;R.\nonumber
\end{equation}
Below we neglect the difference between $\varphi^L$ and
$\varphi^T$, $f^L$ and $f^T$.

It is useful to introduce dimensionless vector ${\bf r}$ via
equation $ {\bf k}_{\bot}= ({\bf r}+{\bf n}) p_{\bot}/2$. Then
the above impact--factors get the forms:
\begin{eqnarray}
J_{\gamma ^* V}& = & eQ _V g^2 {\delta _{ab} \over 2N} \;{ f_V
\over \mid p_{\bot}\mid}
\left\{\begin{array}{ll}
\left({\bf e}{\bf F}_{T\to V_L}\right)& \mbox{ for T-photon}\to
\mbox { meson }V_L\\
\delta\left({\bf e}{\bf e} ^*_V\right) F_{T\to V_T}&
\mbox{ for T-photon}\to\mbox { meson }V_T\\
F_{S\to V_L} & \mbox{ for S-photon}\to\mbox { meson }V_L.
\end{array}\right.\label{12}\\
{\bf F}_{T\to V_L}& =& - \int \limits^{1}_{-1}d\xi \varphi _V(\xi
)\cdot \xi
\left\{\left[\frac{(1+\xi){\bf n}}{v+(1+\xi)^2 }+
\frac{{\bf r}-{\bf n}\xi}{v+\left({\bf r}-{\bf n}\xi\right)^2}
\right]-\left[\xi \leftrightarrow - \xi\right]\right\};\label{12a}\\
F_{T\to V_{T}}&=&\int \limits^{1}_{-1}\varphi_V(\xi) d\xi\cdot
{\cal R};\quad F_{S\to V_{L}}=-\int \limits^{1}_{-1}d\xi
\sqrt{u}(1-\xi^2)\varphi_V(\xi)
\cdot{\cal R}\label{12b}\\
{\cal R}&=& \left[\frac{1}{v+(1+\xi)^2 }-\frac{1}{v+\left({\bf
r}+ {\bf n}\xi\right)^2}\right]+\left[\xi \leftrightarrow -
\xi\right].\nonumber
\end{eqnarray}

{\bf The impact--factor for {\boldmath $\gamma\to \gamma$}
transition} (for the hard Compton effect) with on shell initial
photon was obtained in fact for a long time ago in refs.
\cite{ChWu,LFr} and written in the convenient form in ref.
\cite{GPS1} neglecting quark mass.
\vspace{2cm}

\section {Neutral vector meson photoproduction on a quark or a
gluon}

To calculate the amplitudes under interest we substitute these
impact--factors into eq. (\ref{4}). The result for the meson
production on a quark is\footnote{ The results in the next three
section were presented earlier in preprints \cite{GIv1} -- without
integration over $\xi$, and \cite{GIS2} -- in the final form.}
\begin{equation}
M_{\gamma ^*q\to Vq}=i {eQ_V g^4 \over \pi}\; {s f_V\over
|p_{\bot}| ^3} \;{N^2-1\over N^2}\; \left\{
\begin{array}{ll}
\left({\bf e}{\bf n}\right)I_{T\to V_L}  &\mbox{ for T-photon}
\to\mbox { meson }V_L \\
\left({\bf e}{\bf e}_V^{\ast} \right)
\delta \cdot I_{T\to V_T}  & \mbox{ for T-photon}\to
\mbox { meson }V_T \\
I_{S\to V_L} &\mbox{ for S-photon}\to\mbox { meson }V_L
\end{array}\right.\label{15}
\end{equation}
with
\begin{equation}
I_a\,=\,{1\over 4\pi}
\int \frac{\displaystyle {F_a({\bf r},{\bf n})}}
{\displaystyle {\left({\bf r}-{\bf n}\right)^2\left({\bf r}+ {\bf
n}\right)^2}} d^2r\equiv
\int \limits^{1}_{-1}d\xi\;\varphi_{V}(\xi)\Phi_a(\xi)
\;\;\left(a= T\to V_L,\;T\to V_T,\; S\to V_L\right).\label{T0}
\end{equation}
(For $a= T\to V_L$ the quantity $({\bf n F}_{T\to V_L})$ is
used.)

Just as in refs. \cite{GPS1},\cite{GPS2}, we perform first
integration over the component of vector ${\bf r}$ along ${\bf
n}$ using residues. The last integration is trivial (but
bulky). Then the quantities in eq. (\ref{15}) get the form
\begin{eqnarray}
\Phi_{T\to V_L}&=&{\xi \over 4(1-\xi^2-v )}
\left [{(1+\xi )^2-v \over (1+\xi )^2+v}
\ln{{(1+\xi )^2+v \over 2\sqrt{v}}}  -
{(1-\xi )^2-v \over (1-\xi )^2+v}
\ln{{(1-\xi )^2+v \over 2\sqrt{v}}}
\right ] \nonumber\\
\Phi_{T\to V_T}&=&{1 \over 2(1-\xi^2-v )}
\left [{(1+\xi ) \over (1+\xi )^2+v}
\ln{{(1+\xi )^2+v \over 2\sqrt{v}}}  +
{(1-\xi ) \over (1-\xi )^2+v}
\ln{{(1-\xi )^2+v \over 2\sqrt{v}}}
\right ] \nonumber\\
\Phi_{S\to V_L}&=&-\sqrt{u}(1-\xi^2)\Phi_{T\to V_T}\ .
\label{16}
\end{eqnarray}

We will discuss below the scale of $Q^2$--dependence for the
cross sections. Let us define this scale $\Lambda^2$ by the
relation
\begin{equation}
I_T(\Lambda^2) ={1\over 2}I_T(Q^2=0).
\end{equation}

The results differ for the production of mesons consisting
of heavy quarks (heavy mesons) and light quarks (light mesons)
or photons.

\section{Production of heavy mesons}

We begin with a more simple case of mesons consisting of heavy
quarks ($J/\Psi$ or $\Upsilon$). It seems more clean since the
large quark mass suppresses nonperturbative effects. The results
are similar in main features to those obtained in ref.\cite{GPS2}
for the mass shell photons. We will speak below about the
$J/\Psi$ meson photoproduction for definiteness.

For the wave function of discussed mesons we use the
main (non--relativistic) approximation:
\begin{equation}
\varphi(\xi)\;=\;\delta(\xi).\label{3a}
\end{equation}
With this wave function the impact--factor of $\gamma^T\to V_L$
transition (for production of longitudinally polarized vector
meson by transverse photon) vanishes. The deviation from the
simple form (\ref{3a}) can be described by the quantity
$<\xi^2>=\int\xi^2\varphi(\xi)\, d\xi \sim 0.1$.
\vspace{0.3cm}

{\bf The photoproduction on a quark.}
\vspace{0.3cm}

The main results for transverse photon coincide with those for
real photons \cite{GPS2} with the natural replacement
$\delta^2\to \nu$. Finally, in eq. (\ref{15})\footnote{ The
result for the spin--nonflip amplitude coincides with that in
ref. \cite{ForRys}). }
\begin{eqnarray}
I_{T\to \Psi_T}&=& {1\over 2(\nu^2-1)} L(\nu);\quad L(\nu)=
\ln{{(1+\nu )^2 \over 4\nu } };\;I_{S\to \Psi_L}=-\sqrt{u}
I_{T\to \Psi_T};\label{PsiT}\\ I_{T\to \Psi _L}&=& {<\xi ^2>\over
(1+\nu)^2}\left[1-{\nu
\over \nu-1} L(\nu)\right];\qquad \nu =
(4m^2+Q^2)/{\bf p}_{\bot}^2 .\label{Psi0}
\end{eqnarray}

The helicity conserves in these reactions with high precision.
{\em The transverse photon produces mainly transverse $\Psi$}.
The admixture of longitudinally polarized $\Psi$ is
$\sim(<\xi^2>)^2 \sim 0.01 $ \cite{GPS2}.  {\em The scalar
photon produces longitudinal $\Psi _L$ only.} The ratio of
amplitude with the production of transverse $J/\Psi$ by
$T$-photons to that with the production of longitudinal $J/\Psi$
by $S$-photons is $\sqrt{4m_c^2/ Q^2}\,$, it is independent on
$p_{\bot}$. At $Q^2\;> 4m^2$ the dominant polarization becomes
longitudinal.

The largest amplitudes $\gamma^*_T q\to \Psi_T q$ and $\gamma^*_S
q\to \Psi_L q$ vanish at $p^2_\bot = 4m_c^2+Q^2$ (or $\nu =1$).
These zeroes shift strongly due to $<\xi^2>$ corrections. At the
higher values of $p_{\bot}$ these cross sections are small (cf.
ref. \cite{GPS2}).

The shape of both main amplitudes is determined by the single
function $I_{T\to \Psi_T}$. One can see that the scale of $Q^2$
dependence $\Lambda^2 $ increases from the natural value
$4m_c^2/2$ at small $p_{\bot}$ to $\sim p_{\bot}^2/10$ at large
enough $p_{\bot}$.

The $Q^2$ dependence for the high energy asymptotic of LLA result
in the photoproduction of heavy meson was considered in ref.
\cite{ForRys}. Despite the difference in the absolute values, the
qualitative picture at small $Q^2$ is similar to that in the
two--gluon approximation.
\vspace{0.3cm}

{\bf\boldmath The production of two mesons in $\gamma\gamma\,$
collision}.
\vspace{0.3cm}

We have considered the production of both identical and different
mesons by real or virtual photons. In particular, the collision
of the virtual photon with the real one is described by two
nonzero amplitudes, the first --- for the production by
$T$-photon and the second --- for the production by $S$-photon
(these amplitudes are finite at $p_\bot
\to 0$):
\begin{eqnarray}
M_{\gamma ^*_T \gamma \to V'_T V_T}&=& {is \over p_{\bot} ^4}
\;{e^2 g^4 \over \pi} Q_V Q_{V'} f_V f_{V'} {N^2-1\over N^2}\;
({\bf e}_1 {\bf e}_{V'}^*)\;({\bf e}_2 {\bf e}_V^*)\;\delta
\delta '\cdot I_{V'V} ; \nonumber\\
M_{\gamma ^*_S \gamma \to V'_L V_T }&=& -{is \over p_{\bot} ^4}
\;{e^2 g^4 \over \pi} Q_V Q_{V'} f_V f_{V'} {N^2-1\over N^2}
\;({\bf e}_2 {\bf e} _{V}^*)\;\sqrt{u'} \delta\cdot I_{V'V} ;
\label{V'V}\\
&&I_{V'V}={1\over u'} \left[ {L(\delta^2)\over (\nu'+1)(\delta
-1)}\; - \; {L(\nu')\over (\nu'-1)(\delta +1)}\right].
\nonumber
\end{eqnarray}
Here $\nu'$ corresponds to the meson produced by off shell
photon, and $\delta$ --- to the meson produced by on shell
photon.

\section{Production of photons and light mesons
on a quark or a gluon}

The pQCD limit here corresponds to the asymptotic
$m/p_{\bot}\to 0$. Therefore, we neglect quark mass in this
subsection.

The amplitude for the hard Compton effect on a quark with on
shell initial photon was obtained in fact in refs.
\cite{ChWu,LFr} and written in the suitable form in ref.
\cite{GPS1} neglecting quark mass. In the region of pQCD
validity one can write for unpolarized photons
\begin{equation}
\sigma(\gamma q\to \gamma q)=\left({p_{\bot}\over 16.6\mbox {GeV}}
\right)^2\sigma(\gamma q\to\rho q).\label{Compt}
\end{equation}

For the vector meson production the impact--factor
$J_{\gamma^*_T V_T}$ contains the factor $\delta = 2m/p_{\bot}$.
Therefore, {\em within the range of pQCD validity (at large enough
$p_{\bot} $) transverse photon produces light meson in the
states with helicity 0 only} for any polarization of initial
photon. This fact is in the evident contrast with the well known
helicity conservation at small $p_{\bot}$ (vector mesons are
produced transversely polarized mainly). It is due to the chiral
nature of perturbative couplings in the massless limit.

We write in this section for the shortness
\begin{equation}
I_{T}\equiv I_{T\to V_L};\quad
I_S\equiv I_{S\to V_L}(u)\,\equiv\,- {2\sqrt{u}\over
1+u}(I_{T}\,+\,U).\label{IST}
\end{equation}

We use the wave functions of mesons consisting of light quarks in
the form \cite{ChAZh}:
\begin{equation}
\varphi _V(\xi )=\frac{3}{4} \left( 1-\xi
^2\right)\left(1-\frac{1}{5}b_V+b_V\xi^2\right).\label{3}
\end{equation}
Coefficients $b_{\rho}=b_{\omega}=1.5,\; b_{\phi}=0$ at
$p_{\bot}\approx 1$ GeV, they tend to 0 slowly with
$p_{\bot}$ growth.

For the asymptotical wave function ($b_V = 0$) we obtain:
\begin{eqnarray}
I_T(u)&\equiv& I_0 (u)=\frac{\displaystyle 3}{\displaystyle
8(1-u)^3} \left[ 2+10u-u\left(\frac{\displaystyle
1+u}{\displaystyle 1-u}\right) \left(\ln^2{\frac{\displaystyle
1}{\displaystyle u}}+6\ln{ \frac{\displaystyle 1}{\displaystyle
u}}\right)\right]; \nonumber\\
U(u)&\equiv&U_0(u)\,=\,\frac{\displaystyle 3}{\displaystyle 8(1-u)}
\left( 2- \frac{\displaystyle 1+u}{\displaystyle 1-u}
\ln{\frac{\displaystyle 1}{\displaystyle u}}
\right).\label{19}
\end{eqnarray}
For the case $b_V\neq 0$ the more complicated expressions are
obtained:
\begin{eqnarray}
I_{T}(u)&=&I_0 (u)\left [ 1- {b_V\over 5}+b_V \left( {1+u\over
1-u}\right)^2 \right] + \nonumber\\ &+&{ b_V\over
12(1-u)^4}\left[-(3-10u)(1-u)+16u(u^2+5u+1)U_0
(u)\right];\label{21}\\
U(u)&=&U_0(u)+\,\frac{b_V}{60(1-u)^2}\left[ 5(1-u)
\,+\,8(1+8u+u^2) U_0(u)\right].\nonumber
\end{eqnarray}
These expressions are regular at $u=1$. At $u=0$ (real
photoproduction) $I_T\;=\;{3\over 4}(1+{7\over 15} b_V)$
\cite{GPS1}.

The dependence on the photon virtuality is concentrated in the
factors $I_{T},\; I_{S}$. The shapes of these functions depend
weakly on the form of wave function (value of quantity $b_V$).
They are plotted in Fig. \ref{fig3} for $\rho ^0 $ meson
production ($b_V=1.5$).

The obtained in pQCD cross sections for the light vector
meson production on a gluon (the sum $d(\sigma _{\gamma ^* _T
g\to V g} +d\sigma _{\gamma ^* _S g\to V g})/ dp^2_{\bot}$ ) are
presented in Fig.\ref{fig4}. They are $s$-independent in the used
two-gluon approximation.

{\em If the virtuality of photon is less than $p_{\bot}^2$
($u<1$)}, the amplitude for transverse photon dominates over
longitudinal one. Fig. \ref{fig3} shows very sharp peak in $I_T$
near $Q^2= 0$. It is due to the items $\propto u\ln^2 u$ in eqs.
(\ref{19}), (\ref{21}). The derivative of the amplitude in $Q^2$
(in $u$) diverges at $Q^2=0$, it is infrared unstable in contrast
with the amplitude itself, which is infrared stable. The quantity
$I_T$ is reduced by half at $u\approx 0.1$. It means, that the
scale of $Q^2$ dependence here is
\begin{equation}
\Lambda^2_{pert} \approx p_{\bot}^2/10.\label{lpert}
\end{equation}

The function $I_{S}(u)$ changes its sign at small enough $u=u_0$.
The quantity $u_0\approx 0.1$ for the asymptotical form of wave function
($b_V=0$) and $u_0\approx 0.02$ for the more wide wave function
with $b_V=1.5$. This behavior is similar in some sense to
that for $J/\Psi$ production** at very high values of $p_{\bot}$.

{\bf The obtained scale of $Q^2$ dependence is substantially
lower than it was expected before calculations.} This fact should
be considers at the calculation of the production rate in
$ep$ collisions.

{\em If the photon virtuality is large, $Q^2 > p_{\bot}^2$} (or
$u>1$), the amplitude with the scalar photon become dominant.
\begin{equation}
M_{\gamma ^*_S q\to V_L q} \propto {\ln{u} \over (Q^2)^{3/2}};\quad
M_{\gamma ^*_T q\to V_L q} \propto {p_{\bot} (\ln {u})^2 \over
(Q^2)^{2}}\quad \left(u={Q^2\over p_{\bot}^2}\right).
\label{u>1}
\end{equation}

This  region of parameters was studied in refs.
\cite{ForRys} for the heavy meson photoproduction at the high
energy asymptotic of LLA. Here the LLA amplitude
contains an additional factor $\propto Q/p_{\bot}$ in comparison
with the two--gluon amplitude.  The same factor for the light
meson photoproduction was found in ref. \cite{Iv}.

\section{ Small $p_{\bot}$ limit for high $Q^2\; (u<1)$
and coherence}

In the pQCD limit the amplitudes of photoproduction on a quark or
a gluon (\ref{PsiT},\ref{19},\ref{21}) diverge at $p_{\bot} \to
0$ (see \cite{GPS2} too). It means that "soft" nonperturbative
region of $k_{\bot}\stackrel{<}{\sim}\mu$ contributes
substantially in this case and pQCD calculations are infrared
unstable. Therefore, the details of $p_{\bot}$ dependence at
$p_{\bot} \to 0$ are out of the range of pQCD validity even at
large $Q^2$ when we consider the production on a color target.

The reason for this divergence is simple: in the above limit the
poles of the both gluon propagators coincide, and we deal with
the integral of the form
$$
\int J({\bf k}_{\bot}) d^2k_{\bot}/ k_{\bot}^4.
$$ In accordance with eq. (\ref{dipol}), $J \propto {\bf
k}_{\bot}$ at ${\bf k}_{\bot} \to 0$ and $J \propto({\bf
p-k})_{\bot}$ at $({\bf p-k})_{\bot}\to 0$. At ${\bf p}_{\bot}\to
0$ both these zeroes coincide, and $J \propto k_{\bot}^2$. The
discussed divergence is only logarithmic, the total cross section
is finite.

On the contrary, the $\gamma^*\gamma\to \Psi \Psi$ amplitude is
finite at $p_{\bot}\to 0$ due to additional factor $J\propto
k_{\bot}^2$ in the integrand. Therefore, the soft part of
integration region gives a negligible contribution here.

The main difference between these amplitudes originates from the
fact that in the last case we deal with the collision of two
colorless objects; the coherence between quarks results in the
additional suppression of soft nonperturbative contributions here
(\ref{dipol}).

The above comparison shows us that the coherence in the both
collided particles should be taken into account to describe
phenomena at any $p_{\bot}$ within pQCD even in the region of
large $Q^2$.

\section{The range of validity of the pQCD results. The
light vector meson production and hard Compton effect}

The above results show us that the using of pQCD for the
description of experimental data can be inaccurate in some region
of parameters. For example, the obtained scale of $Q^2$
dependence $\Lambda^2_{pert}\approx p_{\bot}^2/10$ (\ref{lpert})
is smaller than the natural scale of this dependence near mass
shell $\Lambda^2_{soft} \approx m_{\rho}^2$ even at $p_{\bot}
=2.5$ GeV when our small parameter $\mu^2/p_{\bot}^2< 0.02$.

In the discussion below we assume the impact representation to be
valid independent on validity of pQCD for the description of
different factors in it. In particular, the proof of impact
representation in the lowest nontrivial approximation of pQCD is
valid even in the regions near the poles of quark propagators in
the impact--factor, where its perturbative form
(\ref{7},\ref{7a}) becomes incorrect.
\vspace{0.3cm}

{\bf The model for amplitude near the bound of pQCD validity
region}
\vspace{0.3cm}

To study the bound of pQCD validity, we use single scale of QCD
nonperturbative effects $\mu$. We will have in mind the value
$\mu=0.2\div 0.3$ GeV, which is close to the confinement scale,
constituent quark mass, mean transverse momentum of quarks in
meson, etc. The value $\mu=180$ MeV is obtained at the analysis
of total $\gamma\gamma$ cross section in the model with quark
and one-gluon exchange
\cite{GevT}.

We simulate nonperturbative effects by the adding of quantity
$\mu^2$ (instead of $m^2$) in the all quark propagator
denominators.  Besides, we change the quantity $m$ from the
quark propagator numerator (in front of item $R$ in eq.
(\ref{5})) for some new quantity $A\sim\mu$:
\begin{equation}
{\bf P}(m)\to {\bf P}(\mu); \qquad mR(m)\to A\cdot R(\mu)\quad
(A\sim \mu).
\label{amu}
\end{equation}

The regions, where the amplitude is sensitive to $\mu$ value, are
beyond the pQCD validity. {\bf\em We denote the bound ${\bf
p_{pert}}$ of pQCD validity region by the relation}
\begin{equation}
d\sigma(p_{\bot}\geq p_{pert}|\mu) > 0.5\cdot
d\sigma(p_{\bot}\geq p_{pert}|\mu =0).
\label{ppertdef}
\end{equation}

\subsection{ Processes with real incident photon}

The main part of discussion in this section is devoted to the
meson photoproduction.

Contribution of item $R$ provides helicity conservation
(production of $V_T$) in the process. The item {\bf P} in the
impact--factor gives longitudinal polarization of produced mesons
($V_L$). The contribution of this item decreases more slow with
$p_{\bot}$ due to extra degree of momentum in the nominator.
Therefore, this item describes the amplitude at high enough
$p_{\bot}$.

Let us discuss the limit $\mu \ll p_{\bot}$ in more detail. The
contribution of $R$ diverges in this limit, due to the
integration near the poles of quark propagators at {\boldmath
$k_{\bot}=q_{i\bot}$}, it is $\sim \ln(p_{\bot}^2 /\mu^2)$ (i.e.
infrared unstable). It dominates at not too large $p_{\bot}$.
Oppositely, the contribution of ${\bf P}$ is finite in the
discussed limit. It is infrared stable, and it defines the
amplitude within the range of pQCD validity.

Therefore, it is natural to assume that {\bf the contribution
{\boldmath $P$} describes the point--like component of a photon}
in the region where confinement effects are negligible. It
dominates at high values of $p_{\bot}$ and it is responsible for the
production of longitudinally polarized mesons. Similarly, {\bf
\boldmath the contribution $R$ for transverse photons describes
the hadron--like component of a photon}. It dominates at not too
high values of $p_{\bot}$ and it gives helicity
conservation here. In addition to the boundary $p_{pert}$
(\ref{ppertdef}) we denote the boundary value ${\bf p_{hel}}$ by
the condition: At $p_{\bot}> p_{hel}$ the mean helicity of
produced meson changes from the transversal to the longitudinal
one (i.e. the hadron--like component $R$ becomes relatively
small).
\vspace{0.3cm}

{\bf The bound of pQCD validity, the estimate of
$p_{pert}$ for mesons}
\vspace{0.3cm}

We expect that $p_{hel}<p_{pert}$. Therefore, to find the bound
$p_{pert}$, one should consider point--like component of a photon
(contribution {\bf P} in the impact--factor) only. We present two
estimate here.

{\em First estimate}. It is well known that the typical scale of
the $Q^2$ dependence for soft processes $\Lambda_{soft}^2\approx
m_{\rho}^2$ (here $m_{\rho}$ is the mass of $\rho$ meson). The
known data show us that this scale increases with the growth of
$p_{\bot}$.

The scale of $Q^2$ dependence for the $\rho$ photoproduction is
$\Lambda^2_{pert}\approx p_{\bot}^2/10$ (\ref{lpert}). The pQCD
can be valid for description of the discussed phenomena only if
$\Lambda^2_{pert}>\Lambda^2_{soft}$, i.e. at $p_{\bot}^2/ 10\;
>\; m_{\rho}^2$, that leads to $p_{\bot}\gtrsim 3$ GeV. It does
not contradict more refined estimate (\ref{ppert}).

{\em Second estimate.} We calculated numerically the contribution
of item {\bf P} in impact--factor (\ref{15})--(\ref{16}) with
some finite value of $\mu$ for different meson wave functions.
The influence of the nonperturbative effects is described
by the quantity $\Phi(\delta,u) = M(\delta ,u)/ M(\delta =0, u)$.
Naturally, the ratio $\Phi\to 1$ at $(p_{\bot}/\mu)\to
\infty$.(That is the pQCD limit.) The value $p_{pert}$ is
obtained via eq. (\ref{ppertdef}). At higher values of $p_{\bot}$
the influence of confinement effects for the pQCD result in the
cross section is described by factor $\Phi^2$, which is between
0.5 and 1. In Fig. \ref{fig5} we present the quantities
$\Phi(\delta,u) $ (for real photons).

It is seen that for mass shell photons $p_{pert}\approx (30\div
40)\mu$. In this region the coefficient $b_V$ in the $\rho$ meson
wave function decreases up to $b_V\approx 0.7$. Taking this fact
into account, we have for $\mu =0.2$ GeV
\begin{equation}
p_{pert} \approx \left\{
\begin{array}{rlr}
7.5\mbox{ GeV}& \mbox{ for } b_V=0.7& (\rho - \mbox{ meson at }
p_{\bot}\approx 7 \mbox { GeV}),\\
6.2\mbox{ GeV}& \mbox{ for } b_V=0 &(\phi - \mbox{ meson}).
\end{array}\right .\label{ppert}
\end{equation}
For $\mu=0.3$ GeV these quantities should be 1.5 times higher.

The obtained values of $p_{pert}$ (\ref{ppert}) for the mass
shell photons are very high. It is because the correction to the
pQCD result is governed by the parameter
$(\mu^2/p_{\bot}^2)\ln^2( p_{\bot}^2/\mu^2) $ but not the
"natural" parameter $\mu^2/p_{\bot}^2$. The effect of "$\mu$
corrections" in pQCD equations is enhanced near the bounds of
kinematical region, at $\xi\to \pm 1$. Therefore, their influence
is higher for the wave function, which is "shifted" to these
bounds (with $b_V>0$).  In other words, the bounds for pQCD
validity region $p_{pert}$ are lower for the $\phi$ meson
photoproduction ($b_V=0$) in comparison with that for $\rho$
photoproduction ($b_V=1.5$).
\vspace{0.3cm}

{\bf The estimate of $p_{pert}$ for the hard Compton
effect}
\vspace{0.3cm}

The amplitude of the hard Compton effect (\ref{0}) for real
photons with the necessary accuracy is obtained from that
in \cite{ChWu} (with some numerical factors, like \cite{GPS1}).
Using the standard quantity $R_\gamma = 3\sum_{uds}Q^2_q=2$
and the quantity $\epsilon=\mu/p_{\bot}$, one can write:
\begin{eqnarray}
&M_{\gamma q\to \gamma q}=i\frac{16}{27}\alpha\alpha_sR_\gamma
\frac{s}{t}\left[A_{\parallel}({\bf e_1n})({\bf e^*_2n})+
A_{\bot}(({\bf e_1e^*_2})-({\bf (e_1n})({\bf e^*_2n})) \right];&\\
& A_{\parallel}=11.58 + 6\epsilon^2\left[
\frac{8}{3}\ln^3\epsilon+2\ln^2\epsilon
-32.32\ln\epsilon-74.83 \right]; & \nonumber\\
& A_{\bot}=5.58 + 6\epsilon^2\left[
6\ln^2\epsilon-10\ln\epsilon+6.6 \right]. &\nonumber
\end{eqnarray}
Two independent amplitudes contribute here. The nonperturbative
corrections in these amplitudes are of different sign. The
largest among them tends faster to its asymptotical value. Naturally,
the bound of the region of the pQCD validity for the
unpolarized photons is relatively low. Finally, using eq.
(\ref{ppertdef}), we obtain:
\begin{eqnarray}
& p_{pert}(A_{\bot})\approx 3 - 4 \mbox { GeV for }
\mu =0.2 - 0.3\mbox { GeV} & \nonumber\\
& p_{pert}(A_{\parallel})\approx 1.3 - 2 \mbox { GeV for }
\mu =0.2 - 0.3\mbox { GeV} &\label{Comptp}\\
& p_{pert}(unpol)\approx 0.8 - 1.2 \mbox { GeV for }
\mu =0.2 - 0.3\mbox { GeV} & \nonumber\\
\end{eqnarray}

Therefore, despite of smaller value of cross section, the study
of pure pQCD behavior in the hard Compton effect can be better
than that in the meson photoproduction.
\vspace{0.3cm}

{\bf Estimate of $p_{hel}$}
\vspace{0.3cm}

Let us find the crossover point, in which the longitudinal
polarization becomes dominant for the transversal initial photon
(the boundary $p_{hel}$). This boundary is less than the boundary
$p_{pert}$. Therefore, the calculations near this point depend on
the details of the model more strong. To see qualitative features
of this crossover, the model (\ref{amu}) is used with the value
$A=1$ GeV.

Figs. \ref{fig6a},\ref{fig6b} shows the cross sections of the
photoproduction by real photons for $\mu=200$ MeV. In these
figures the curves R correspond to the production of transverse
mesons (helicity conserved contribution, item $R$ (\ref{7a}),
hadron--like component of a photon) and the curves P correspond
to the production of longitudinal mesons (item ${\bf P}$
(\ref{7}), point--like component of a photon). Fig. \ref{fig6a}
shows the curves for $\rho^0$ meson production ($b_V=1.5$). Fig.
\ref{fig6b} shows the curves for $\phi$ meson production
(asymptotical wave function, $b_V=0$).

In both cases the crossover point $p_{hel}$ is $1.5\div 5$ GeV.
Next, the admixture of transversely polarized mesons at $
p_{\bot}>p_{hel}$ for the $\rho$ photoproduction decreases
with $p_{\bot}$ faster than that one for $\phi$
mesons. We expect, that this feature conserves for virtual
photons. It means, that in the data averaged over some
$p_{\bot}$ interval the fraction of longitudinal
$\phi$'s is larger than that for $\rho$'s. This conclusion is
supported by the data \cite{exp2}.

\subsection{Processes with the off shell photons}

The photon virtuality prevents quark propagators from their poles
while $\xi\neq\pm 1$. It is the reason why $p_{pert}$ decreases
fast with the photon's virtuality. Our calculations show that
\begin{eqnarray}
p_{pert}^2& \approx& \left\{
\begin{array}{rr}
1.3 \mbox{ GeV}^2& \mbox { for } \rho \\ 1 \mbox{ GeV}^2& \mbox {
for } \phi \\
\end{array}\right\} \mbox { at } \mu =0.2\mbox{ GeV}, Q^2 =1
\mbox{ GeV}^2; \nonumber\\
p_{pert}^2& \approx& \left\{
\begin{array}{rr}
28 \mbox{ GeV}^2& \mbox { for } \rho \\ 10 \mbox{ GeV}^2& \mbox {
for } \phi \\
\end{array}\right\} \mbox { at } \mu =0.3\mbox{ GeV}, Q^2 =1
\mbox{ GeV}^2; \label{ppertq}\\
p_{pert}^2& \approx& \left\{
\begin{array}{rr}
3.3 \mbox{ GeV}^2& \mbox { for } \rho \\ 2 \mbox{ GeV}^2& \mbox {
for } \phi \\
\end{array}\right\} \mbox { at } \mu =0.3\mbox{ GeV}, Q^2 =2.25
\mbox{ GeV}^2; \nonumber\\
\end{eqnarray}

It is seen, that the pure pQCD description for the longitudinal
meson photoproduction becomes valid earlier for the $\phi$
photoproduction as compare with $\rho$ one.\\??

\subsection{ The signature of pQCD validity. }

{\em In the range of pQCD validity the light mesons should be
produced in the state with helicity 0 only} \cite{GPS1}. It is
in strong contrast with the production in "soft" region where
the helicity conservation takes place, and real photons produce
transversely polarized mesons. Unfortunately, there is no
similar good signature for the hard Compton effect and
production of heavy mesons.

Besides, the striking feature of the results obtained is very
{\em sharp dependence on the photon virtuality} of the amplitude
of reactions (\ref{0}) near the point $Q^2=0$ (more precise, on
the ratio $u$). The observation of such behavior will be a good
test of pQCD.

\section{Brief discussion about some related papers}

In the paper \cite{ForRys} the photoproduction of heavy mesons
in the process (\ref{0}) was studied as in the two-gluon
approximation as well as in the LLA (using the BFKL equation) in
dependence on both $p_{\bot}^2$ and $Q^2$. It was found the lack
of validity of the two--gluon approximation when $\rho \ge 0.8$
(see eq. (\ref{BFKLpom}) for description and discussion). At
$\rho\sim 2$ the LLA result exceeds the Born one by a factor
$\sim 10$ in cross section.

In this paper the non--relativistic approximation for the meson
wave function (\ref{3a}) was used.   The calculations in both this
paper and in refs. \cite{Iv,Iv1} show that many features
of results in ref. \cite{ForRys} are connected with this specific
form of the wave function and difference in the quark masses.

To describe the production of light mesons it is necessary to
consider the nontrivial longitudinal motion of quarks in
a light meson according to the standard leading twist
approach to the hard exclusive processes, for example, in the
form (\ref{3}) \cite{LeBr,ChAZh,BaGr}.  This very difference
together with the difference in quark masses results in the
discussed substantial difference in the produced mesons
polarization in the reactions \ref{0}). It is the first point,
which shows that the experience with the $J/\Psi$ photoproduction
is almost useless for the description of the light vector meson
photoproduction.

The light vector mesons production in the process (\ref{0}) was
studied in the high energy limit of LLA in ref. \cite{Iv}.  The
obtained cross section was found to be less than the that in the
two--gluon approximation at $\rho<3.4$ (cf. eq.
(\ref{BFKLpom}).  Even at $c_1,\, c_2=1,\; p_{\bot}=5 $ GeV this
value of $\rho$ corresponds to $\sqrt{s_{\gamma q}}= 335\mbox{
GeV}$. This energy is unaccesible for the HERA collider. For the
smaller values of $p_\bot$ we come to the lower values of
$\sqrt{s_{\gamma q}}$, at which the BFKL result for the
amplitude exceeds the Born one, are lower. However, as it is
shown above, this region is definitely beyond the pQCD validity.
The pure pQCD description based on the "hard Pomeron" concept is
invalid here.  The hard Compton effect is more favorable to
study the pP due to both much lower value of $p_{pert}$
(\ref{Comptp}) and lower value of $\rho$, which is necessary to
see BFKL effect in comparison with the two--gluon approximation
\cite{Iv1}.

Other known for us papers, which treat the similar problems
for the processes like (\ref{0}), contain the essential
phenomenological components (usually pQCD inspired).

These models are based, in fact, on the impact representation
like (\ref{4}). The description begins with the diffractive
region (small $p_{\bot}$). It is a reasons why authors
use the hadron--like component of photon (item $R$ but no
point--like one ${\bf P}$) only with some parameter $\mu$ for the
detail description of cross section.  These models predict the
helicity conservation in reaction $\gamma q\to\rho^0 q$. They do
not predict the change of polarization of produced mesons at high
$p_{\bot}$.

The quasi--elastic process $\gamma^*p\to\rho p$ (without
proton's dissociation) was studied in refs. \cite{Land,Cud}. In
these papers it was used the QCD inspired phenomenological
model, which can be written in the form of impact representation
(\ref{4}) with the replacement of pQCD gluon propagators on the
reggeized ones. The $Q^2$ dependence for the forward scattering
amplitude in this model differs from that obtained in pQCD
\cite{BrF}.

Recently the papers \cite{BrF},\cite{Rys},\cite{Nic} were
published, where the problems are studied that are close to those
discussed above. In these papers quasi--elastic photoproduction of
vector mesons on a proton without proton's dissociation
($\gamma^* p\to Vp$) is studied (with $V=\rho $ in \cite{BrF,Nic}
and $V=J/\Psi$ in \cite{Rys}).

The first stage in these papers corresponds to the simplest pQCD
diagram just as in our paper. The following stages are used some
features of processes at $p_{\bot}\approx 0$. To describe the
picture at large $p_{\bot}$ some phenomenological assumptions
were added in papers \cite{Rys},\cite{Nic}.

The $\rho$ meson photoproduction at $p_{\bot}\approx 0$ was
studied in ref. \cite{BrF}. The same very region for the $J/\Psi$
photoproduction is the starting point for ref. \cite{Rys}. Just
here some features of LLA provides an opportunity to use
unitarity for the construction of cross section in terms of the
LLA proton's gluon distribution\footnote{ This basic construction
is broken up at $p_{\bot}\neq 0$.  To describe the $J/\Psi$
production in this region, it is used the additional assumption
in ref. \cite{Rys}, that the object, which was the proton's gluon
distribution at $p_{\bot}=0$, transforms to the product of this
distribution and some proton form factor dependent on
$p_{\bot}^2$ only.}. We consider other kinematical region,
see eq.~(\ref{range}).

The papers \cite{Nic} treat the process (\ref{0}). The crucial
point here is the using of hadron--like component of a photon
(factor $R$). It is the reason why these authors obtain the
transversal polarization of $\rho$ meson for on shell
photoproduction even at large enough $p_{\bot}$.

\section{ Concluding remarks}

Let us summarize our predictions (mainly for HERA experiments)
related to the photoproduction of light mesons and hard Compton
effect. (We remind that the results for the $J/\Psi$
photoproduction relate weakly to these problems.)

\begin{enumerate}
\item For the real photoproduction we expect the change of mean
polarization of produced vector mesons at $p_{\bot} \sim 1.5\div
5$ GeV. Above this bound the produced vector mesons should be
mainly longitudinally polarized. We expect that for the $\phi$
photoproduction this bound is lower than that for $\rho$, and the
fraction of transversal $\phi$ decreases with $p_{\bot}$ more
fast.

\item The pure pQCD regime is hardly observable for the production
of light mesons by real photons, since the corresponding boundary
value of transverse momentum is very high, even for $\mu =0.2$
GeV (\ref{ppert}). However it seems observable at the study
of hard Compton effect.

We believe that this pQCD regime is observable at HERA and at
future Photon colliders for both light vector meson and photon
production.  For the vector meson production this regime can be
seen better in the photoproduction by virtual photons
(\ref{ppertq}). The signatures for this regime is: {\em The
produced light mesons are polarized longitudinally.}
Unfortunately this pQCD regime seems to be unobsrvable at LEP2.

\item One can consider the special region of large enough
$p_{\bot}$ (within the region of pQCD validity) and not too high
values of rapidity gap (say, $y<3$). In accordance with the
results of refs. \cite{Rys},\cite{Iv}, we expect that in this
region our two gluon approximation works, i.e. the
$y$--dependence is weak and $u$--dependence is given by eqs.
(\ref{19},\ref{21}).
\end{enumerate}

The photoproduction of jets in the "direct" configuration and
with the rapidity gap provides the opportunity to see the same
mechanisms in processes with the larger cross sections. First
data on these processes were reported recently \cite{exp3}. The
results of corresponding calculations are bulky and needs
for detail discussions. One can expect that the point $p_{pert}$
will be lower here than that for the vector meson production
(\ref{ppert}).

Last, the photoproduction of scalar (or tensor) meson on a
gluon is forbidden due to C parity conservation \cite{Gin}.
Therefore, the comparative study of the vector and scalar (or
tensor) meson production in $ep$ collision can give an additional
information about the gluon content of a proton and about the
shadowing effects at small $x$.\\

{\bf Acknowledgments.} This paper was prepared in the continuous
discussions with V.G.~Serbo, who checked some results. We
are very thankful for his help and criticism. We are grateful to
P.~Aurhence, A.C.~Bawa, W.~Buchmuller, V.~Chernyak, R.~Cudell,
A.~Efremov, L.~Lipatov, K.~Melnikov, M.~Ryskin, A.~Vainshtein and
G.~Wolf for useful discussions. This work is supported by the
grants of Russian Foundation of Fundamental Investigations
and INTAS. I.F.G. is thankful to the Soros Educational Program
for support.

\begin{figure}[hbt]
\begin{center}
\unitlength 1mm
\begin{picture}(160,130)(0,-10)
\multiput(36,100)(4,0){9}{\oval(2,1.5)[b]}
\multiput(38,100)(4,0){9}{\oval(2,1.5)[t]}
\put(71,100){\line(2,1){12}}
\put(71,100){\line(2,-1){12}}
\put(96,100){\oval(30,15)}
\put(109,106){\vector(1,0){10}}
\put(125,94){\vector(-1,0){16}}
\put(119,106){\line(1,0){6}}
\put(114,108){\makebox(0,0){$ q_1$}}
\put(114,97){\makebox(0,0){$ q_2$}}
\put(53,104){\makebox(0,0){$ p_1$}}
\put(143,104){\makebox(0,0){$ p'_1=q_1+q_2$}}
\put(135,100){\line(-5,3){10}}
\put(135,100){\line(-5,-3){10}}
\put(134,100.5){\vector(1,0){16}}
\put(134,99.5){\vector(1,0){16}}
\put(35,65){\vector(1,0){115}}
\put(60,65){\vector(1,0){10}}
\multiput(91,67)(0,5){5}{\line(0,1){3}}
\multiput(101,67)(0,5){5}{\line(0,1){3}}
\put(53,68){\makebox(0,0){$ p_2$}}
\put(143,68){\makebox(0,0){$ p_{2}^{'}$}}

\put(88,75){\makebox(0,0){$ k$}}
\put(107,75){\makebox(0,0){$ p-k$}}
\put(130,78){\makebox(0,0){$ p=p'_1-p_1$}}
\put(45,78){\makebox(0,0){$ s=(p_1+p_2)^2$}}

\put(85,61){\makebox(0,0){ \bf A. Basic diagram}}
\multiput(51,45)(4,0){3}{\oval(2,1.5)[b]}
\multiput(53,45)(4,0){3}{\oval(2,1.5)[t]}
\put(62,45){\line(2,1){12}}
\put(62,45){\line(2,-1){12}}
\put(87,45){\oval(30,15)}
\put(100,51){\vector(1,0){10}}
\put(110,39){\vector(-1,0){10}}
\put(105,53){\makebox(0,0){$ q_1$}}
\put(105,42){\makebox(0,0){$ q_2$}}
\put(79,32){\makebox(0,0){$ k$}}
\put(98,32){\makebox(0,0){$ p-k$}}
\put(82,37){\line(0,-1){3}}
\put(82,32){\vector(0,-1){3}}
\put(92,37){\line(0,-1){3}}
\put(92,32){\vector(0,-1){3}}
\put(114,45){\makebox(0,0){$ =$}}

\put(1,19){\oval(2,1.5)[b]}
\put(3,19){\oval(2,1.5)[t]}
\put(4,19){\line(2,1){12}}
\put(4,19){\line(2,-1){12}}
\put(16,25){\vector(1,0){24}}
\put(40,13){\vector(-1,0){14}}
\put(16,13){\line(1,0){10}}
\put(35,27){\makebox(0,0){$ q_1$}}
\put(35,16){\makebox(0,0){$ q_2$}}
\multiput(20,25)(0,-5){4}{\line(0,-1){3}}
\multiput(30,13)(0,-5){2}{\line(0,-1){3}}
\put(18,8){\makebox(0,0){$ k$}}
\put(36,8){\makebox(0,0){$ p-k$}}
\put(20,25){\circle*{1}}
\put(30,13){\circle*{1}}
\put(41,19){\makebox(0,0){$ +$}}

\put(47,19){\oval(2,1.5)[b]}
\put(49,19){\oval(2,1.5)[t]}
\put(50,19){\line(2,1){12}}
\put(50,19){\line(2,-1){12}}
\put(62,25){\vector(1,0){24}}
\put(86,13){\vector(-1,0){14}}
\put(62,13){\line(1,0){10}}
\put(81,27){\makebox(0,0){$ q_1$}}
\put(81,16){\makebox(0,0){$ q_2$}}
\multiput(76,25)(0,-5){4}{\line(0,-1){3}}
\multiput(66,25)(0,-5){4}{\line(0,-1){3}}
\put(64,8){\makebox(0,0){$ k$}}
\put(82,8){\makebox(0,0){$ p-k$}}
\put(76,25){\circle*{1}}
\put(66,25){\circle*{1}}
\put(89,19){\makebox(0,0){$ +$}}

\put(94,19){\oval(2,1.5)[b]}
\put(96,19){\oval(2,1.5)[t]}
\put(97,19){\line(2,1){12}}
\put(97,19){\line(2,-1){12}}
\put(109,25){\vector(1,0){22}}
\put(131,13){\vector(-1,0){14}}
\put(109,13){\line(1,0){8}}
\put(127,27){\makebox(0,0){$ q_1$}}
\put(127,16){\makebox(0,0){$ q_2$}}
\multiput(123,13)(0,-5){2}{\line(0,-1){3}}
\multiput(113,13)(0,-5){2}{\line(0,-1){3}}
\put(111,8){\makebox(0,0){$ k$}}
\put(129,8){\makebox(0,0){$ p-k$}}
\put(123,13){\circle*{1}}
\put(113,13){\circle*{1}}

\put(145,19){\makebox(0,0){\large$+\left( k \leftrightarrow
p-k\right). $}}

\put(80,-5){\makebox(0,0){\bf B. Impact--factor}}

\end{picture}

\end{center}
\caption{Photoproduction of vector meson on a quark
(two--gluon exchange).}

\label{fig1}
\end{figure}

\newpage

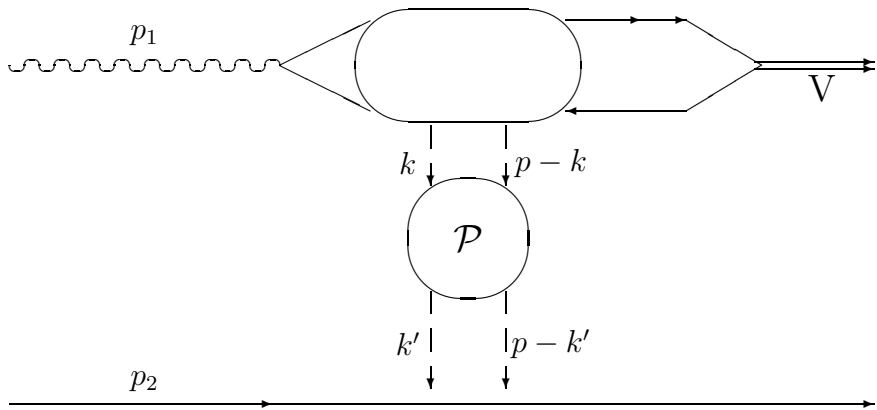
\begin{figure}[hbt]
\begin{center}
\unitlength 1mm
\begin{picture}(120,50)
\multiput(1,45)(4,0){9}{\oval(2,1.5)[b]}
\multiput(3,45)(4,0){9}{\oval(2,1.5)[t]}
\put(36,45){\line(2,1){12}}
\put(36,45){\line(2,-1){12}}
\put(61,45){\oval(30,15)}
\put(74,51){\vector(1,0){10}}
\put(84,51){\vector(1,0){6}}
\put(90,39){\vector(-1,0){16}}
\put(0,0){\vector(1,0){115}}
\put(25,0){\vector(1,0){10}}
\put(18,49){\makebox(0,0){$ p_1$}}
\put(18,3){\makebox(0,0){$ p_2$}}

\put(53,32){\makebox(0,0){$ k$}}
\put(72,32){\makebox(0,0){$ p-k$}}
\put(56,37){\line(0,-1){3}}
\put(56,32){\vector(0,-1){3}}
\multiput(56,7)(0,5){2}{\line(0,1){3}}
\put(56,5){\vector(0,-1){3}}
\put(53,8){\makebox(0,0){$k'$}}
\put(66,37){\line(0,-1){3}}
\put(66,32){\vector(0,-1){3}}
\multiput(66,7)(0,5){2}{\line(0,1){3}}
\put(66,5){\vector(0,-1){3}}
\put(72,8){\makebox(0,0){$ p-k'$}}
\put(61,22){\oval(16,16)}
\put(61,22){\makebox(0,0){\large$\cal P$}}
\put(100,45){\line(-5,3){10}}
\put(100,45){\line(-5,-3){10}}
\put(99,45.5){\vector(1,0){16}}
\put(99,44.5){\vector(1,0){16}}
\put(108,42){\makebox(0,0){\large V}}

\end{picture}

\caption{Pomeron exchange in the process of vector meson photoproduction.}

\end{center}
\label{fig2}

\end{figure}

\begin{figure}
\epsfxsize=15cm
\centerline{\epsffile{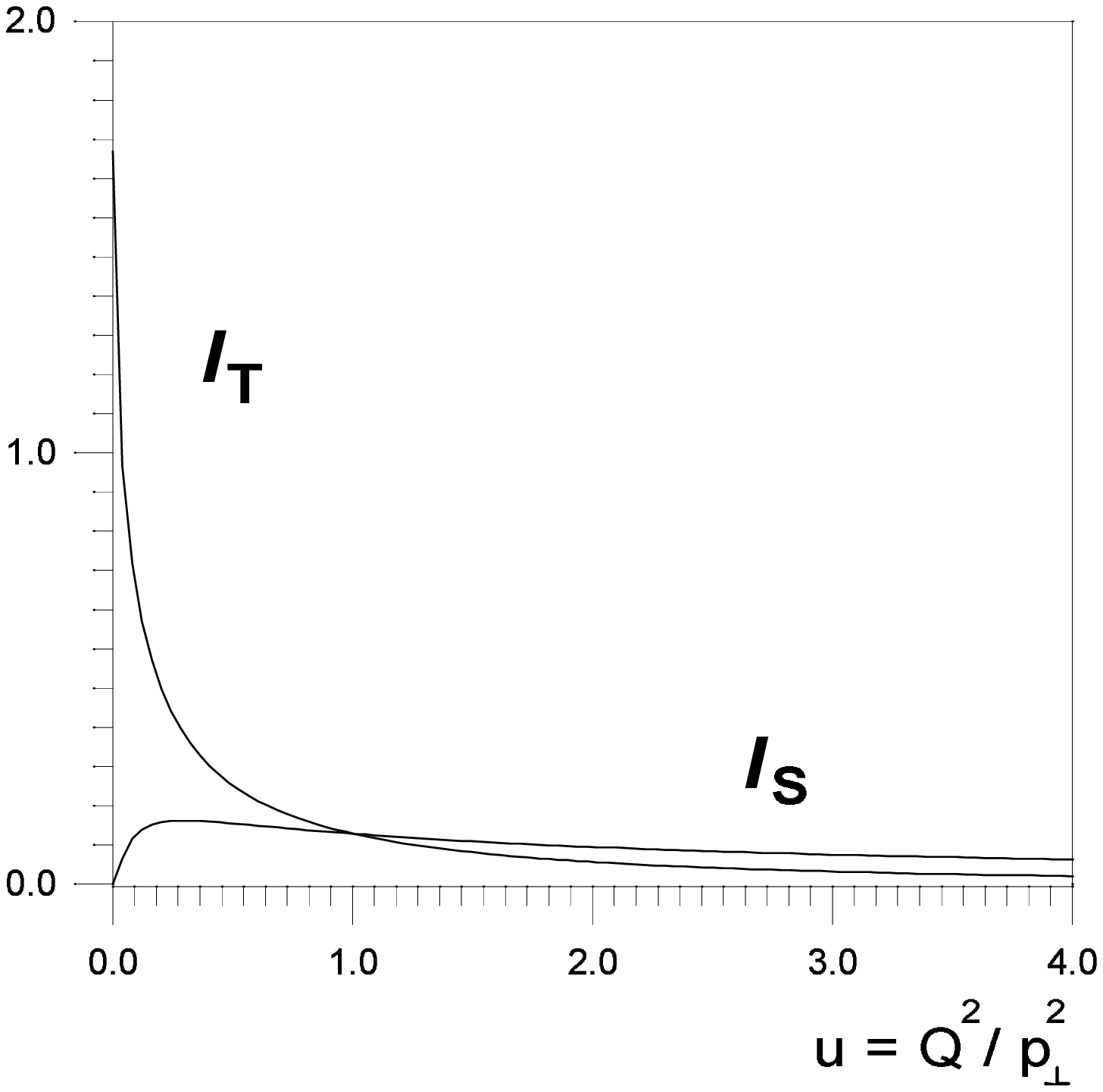}}
\vspace{-1.5cm}
\caption{Functions $I_T$ and $I_S$ for the process $\gamma ^*
q\to \rho^0 q$ or $\gamma ^* g\to \rho ^0 g$.}
\label{fig3}
\end{figure}

\begin{figure}
\epsfxsize=15cm
\centerline{\epsffile{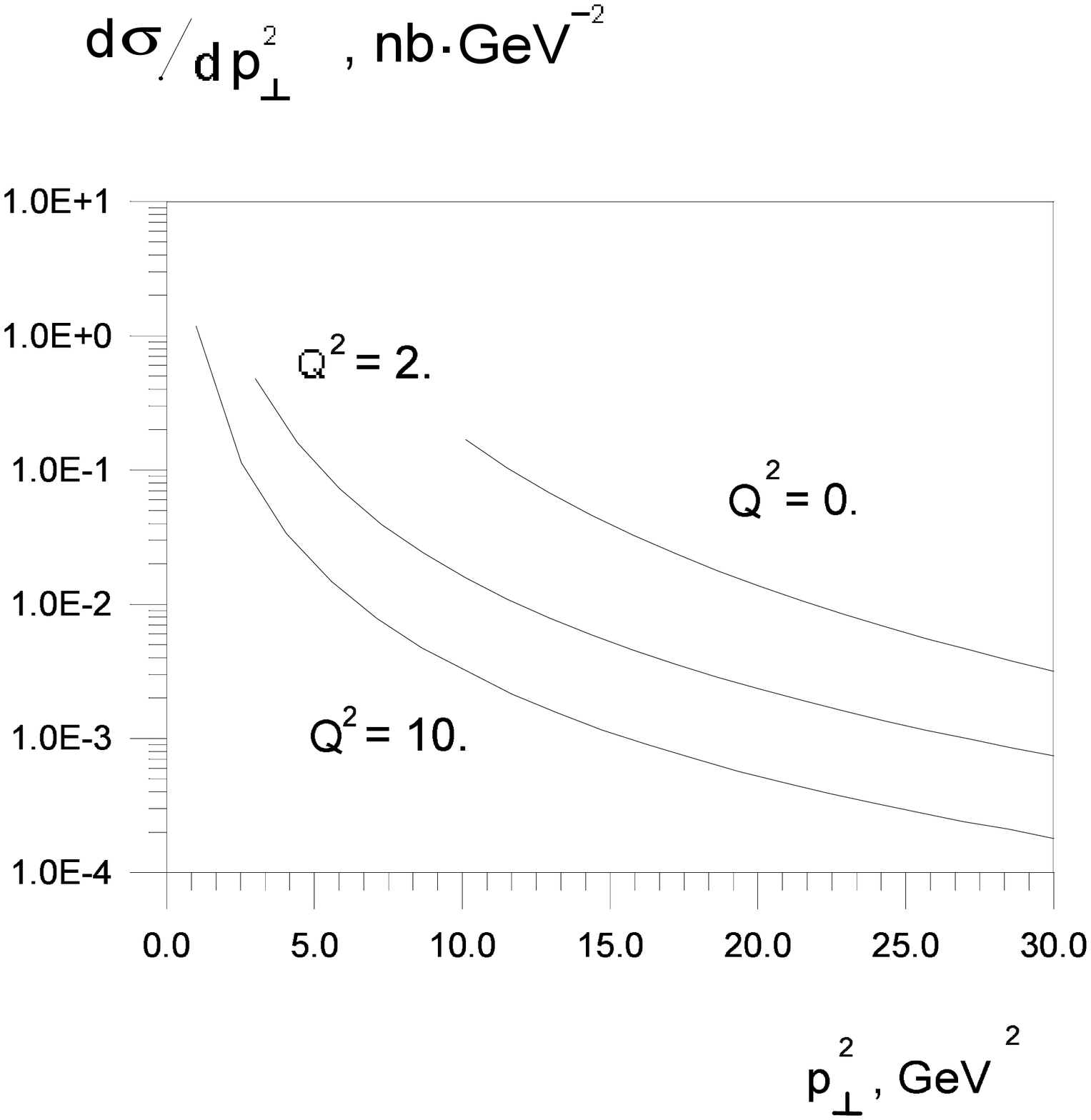}}
\vspace{-1.5cm}
\caption{Differential cross section of $\gamma^* g\to \rho^0 g$
process at $Q^2=0$, $Q^2=2\mbox{ GeV}^2$ and $Q^2= 10\mbox{
GeV}^2$.}
\label{fig4}
\end{figure}

\begin{figure}
\epsfxsize=15cm
\centerline{\epsffile{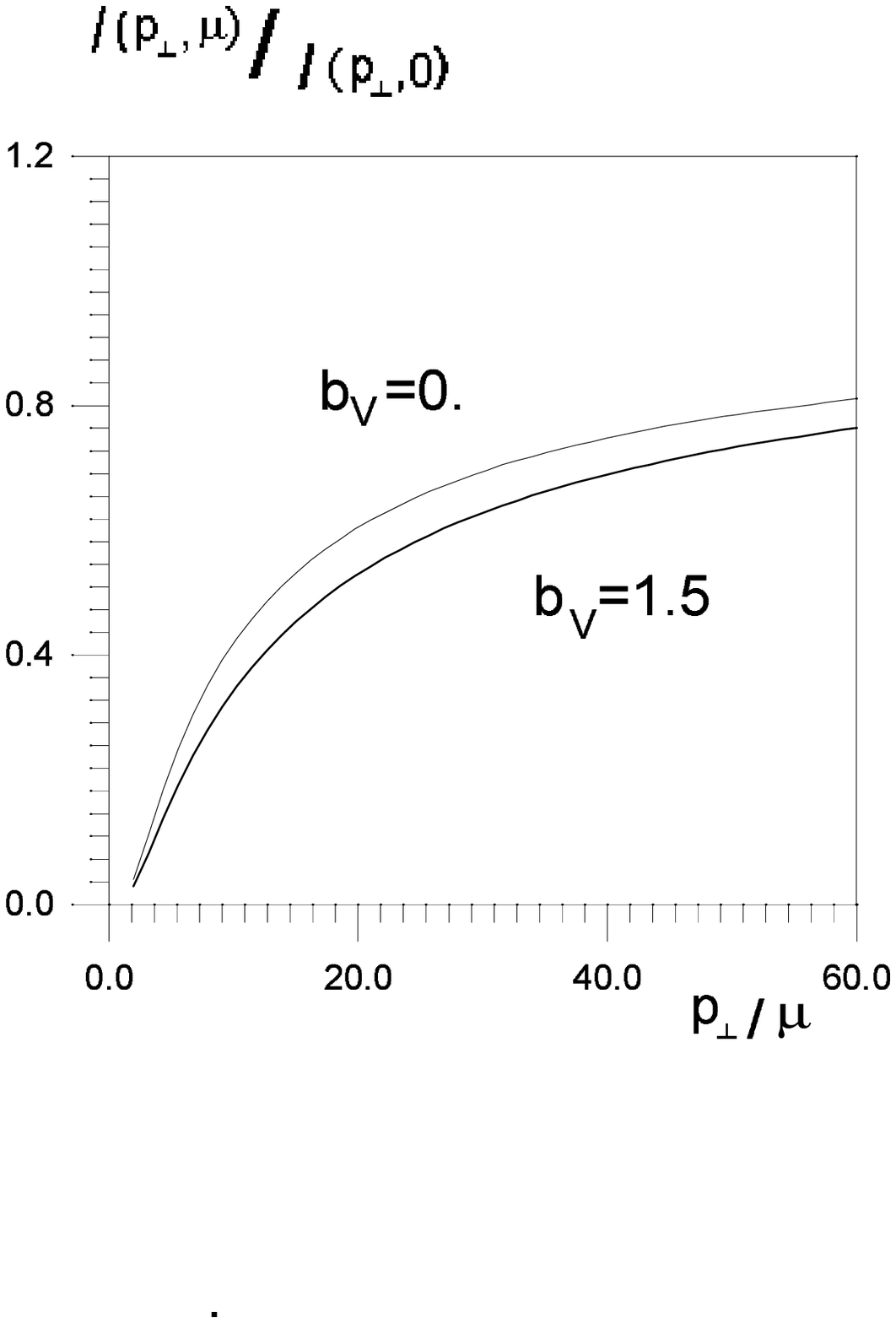}}
\vspace{-1.5cm}
\caption{The ratios $\Phi=M(\delta)/M(\delta =0)$ for mass shell
photons ($u=0$) in dependence on
$p_{\bot}/\mu=2\delta^{-1}$.}
\label{fig5}
\end{figure}

\begin{figure}
\epsfxsize=15cm
\centerline{\epsffile{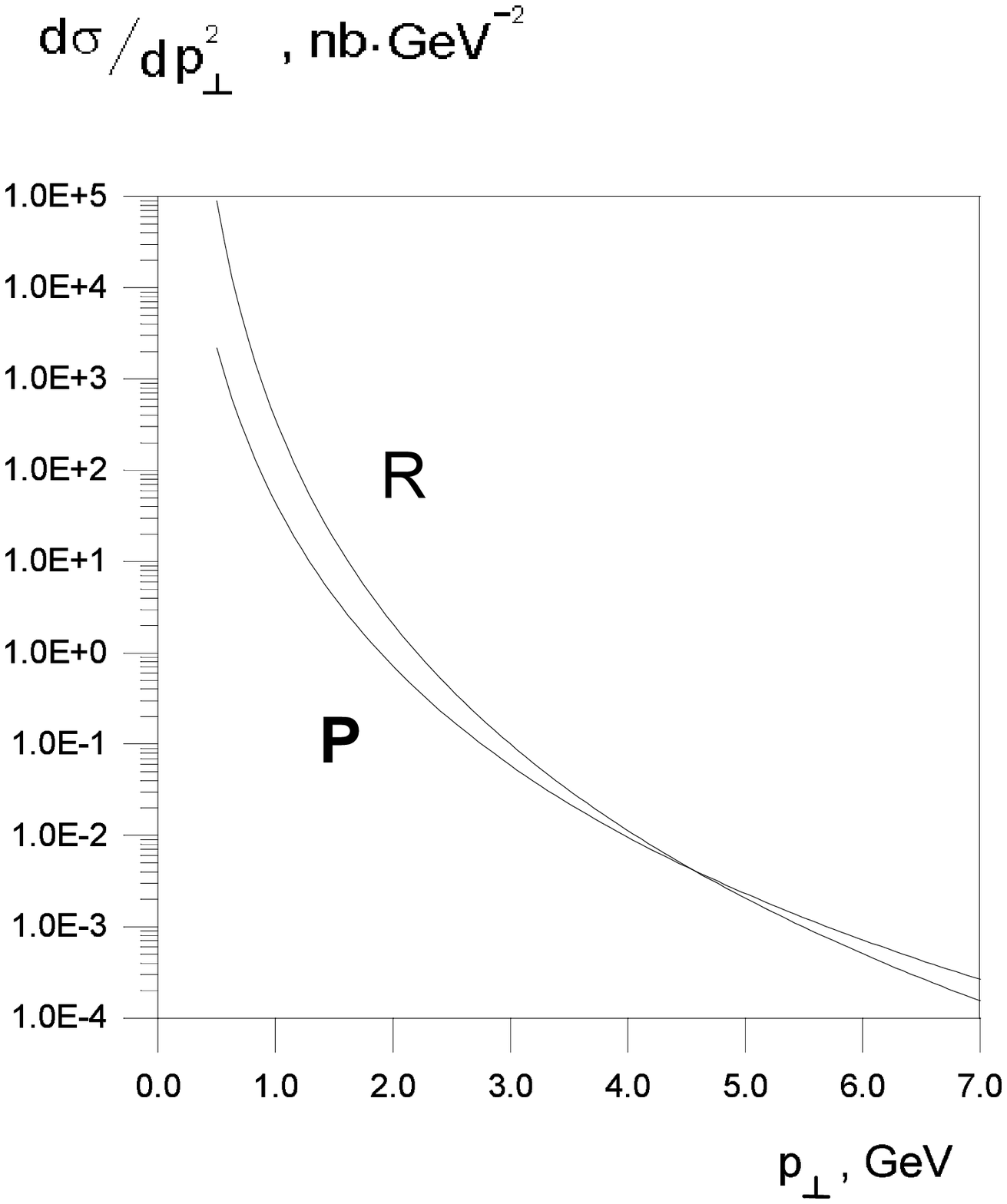}}
\vspace{-1.5cm}
\caption{The differential cross sections of $\rho^0$ meson
photoproduction ($b_V=1.5$) by real photons for $\mu=200$ MeV. In
these figures curves R correspond to the production of transverse
mesons (helicity conserved contribution, item $R$, hadron--like
component of photon) and curves P correspond to the production of
longitudinal mesons (item ${\bf P}$, point--like component of
photon).}
\label{fig6a}
\end{figure}

\begin{figure}
\epsfxsize=15cm
\centerline{\epsffile{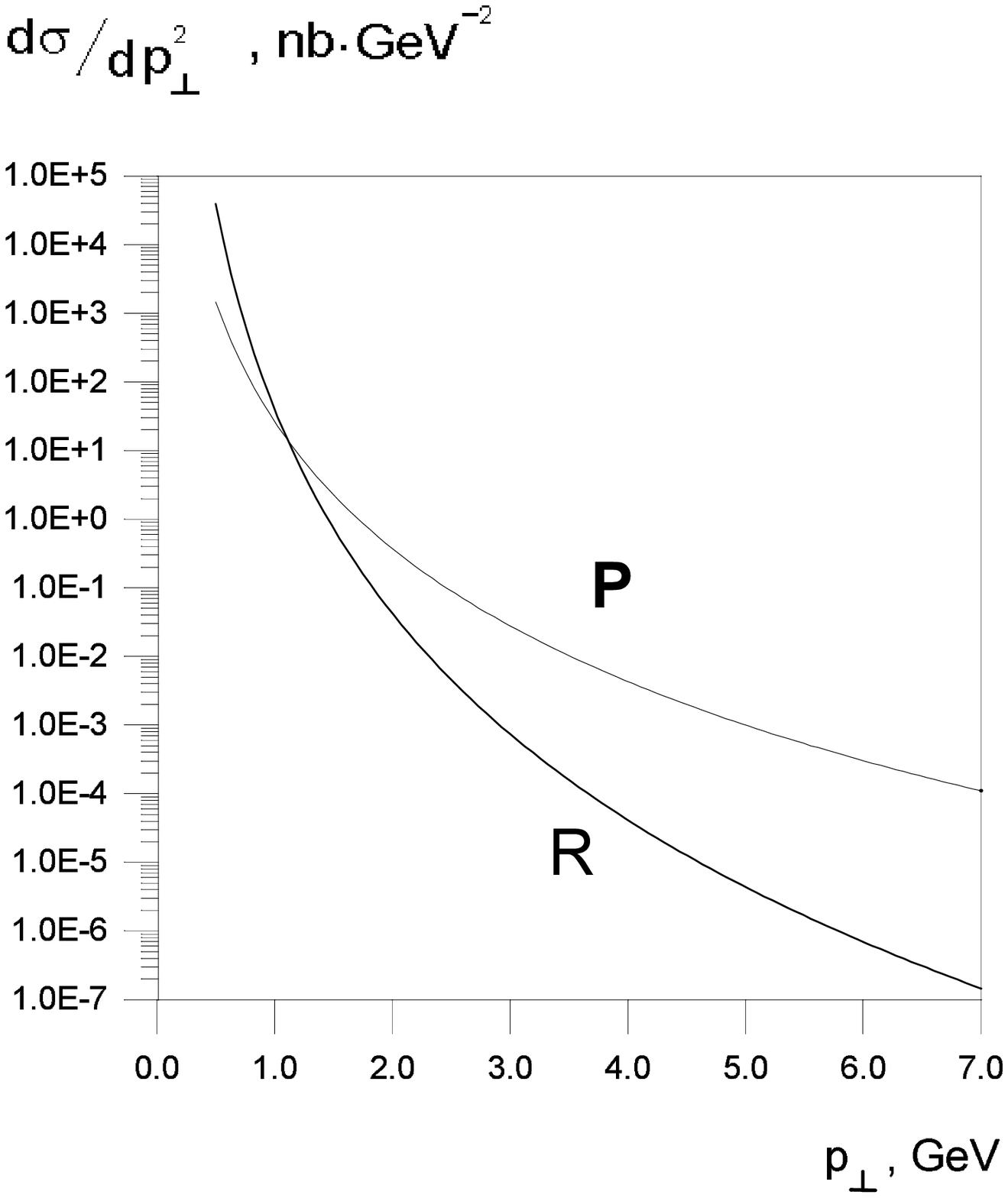}}
\vspace{-1.5cm}
\caption{The same figure as previous one, but for photoproduction
of $\phi$ meson ($b_V=0$). }
\label{fig6b}
\end{figure}

\begin{table}
\caption{Values of mesons coupling constants.}
\label{tab1}

\begin{tabular}{|c|c|c|c|c|c|c|c|c|}\hline
&$\rho^0$&$\omega$&$\phi$&$\Psi$&
$\Psi'$&$\Upsilon$&$\Upsilon'$&$\Upsilon''$\\ \hline
&&&&&&&&\\$f_V$, GeV &0.21&0.21&0.23&0.38&0.28&0.66&0.49&0.42\\
&&&&&&&&\\ $Q_V$ &$1/\sqrt {2}$&$1/(3 \sqrt{2})$
&1/3&2/3&2/3&1/3&1/3&1/3\\
\hline
\end{tabular}

\end{table}

\end{document}